# A Practical System for Guaranteed Access in the Presence of DDoS Attacks and Flash Crowds

Yi-Hsuan Kung[‖], Taeho Lee[†], Po-Ning Tseng[‖], Hsu-Chun Hsiao[‖*], Tiffany Hyun-Jin Kim[‡*], Soo Bum Lee[§*], Yue-Hsun Lin[¶*] and Adrian Perrig[†*]
[‖] *National Taiwan University, {b00902071, hchsiao, b00902067}@csie.ntu.edu.tw*
[†] *ETH Zurich, {kthlee, adrian.perrig}@inf.ethz.ch*
[‡] *HRL Laboratories, hjkim@hrl.com*
[§] *Qualcomm, soobuml@qti.qualcomm.com*
[¶] *Samsung Research America, yuehhsun.lin@samsung.com*

***Abstract*—With the growing incidents of flash crowds and sophisticated DDoS attacks mimicking benign traffic, it becomes challenging to protect Internet-based services solely by differentiating attack traffic from legitimate traffic. While fair-sharing schemes are commonly suggested as a defense when differentiation is difficult, they alone may suffer from highly variable or even unbounded waiting times. We propose RainCheck Filter (RCF), a lightweight primitive that guarantees bounded waiting time for clients despite server flooding without keeping per-client state on the server. RCF achieves strong waiting time guarantees by prioritizing clients based on how long the clients have waited—as if the server maintained a queue in which the clients lined up waiting for service. To avoid keeping state for every incoming client request, the server sends to the client a *raincheck*, a timestamped cryptographic token that not only informs the client to retry later but also serves as a proof of the client's priority level within the virtual queue. We prove that every client complying with RCF can access the server in bounded time, even under a flash crowd incident or a DDoS attack. Our large-scale simulations confirm that RCF provides a small and predictable maximum waiting time while existing schemes cannot. To demonstrate its deployability, we implement RCF as a Python module such that web developers can protect a critical server resource by adding only three lines of code.**

## I. Introduction

Internet users are impatient. A recent study found that more than half of online shoppers abandon websites that fail to load in three seconds [1]. When a wait is unavoidable, users perceive known, finite waits to be shorter than uncertain waits [2], and are willing to wait much longer periods given visual feedback, such as a progress bar [3], [4].

In the presence of Distributed Denial of Service (DDoS) attacks, users would suffer from an uncertain or even infinite waiting time for accessing an online service, as neither the server nor the user knows when the attack will cease. Unfortunately, DDoS attacks are easy to launch and can cause severe damage. Enterprise solutions for DDoS protection (e.g., adding more servers or using Content Delivery Networks) may be too costly for small- and medium-sized companies to afford [5], and CDN web hosting may not be suitable for security-sensitive services. Also, sophisticated DDoS attacks successfully emulate flash crowds and become stealthier as the attacks target scarce server resources, such as CPU and disk I/O, with only a low traffic volume [6]. It is therefore difficult, if not impossible, to differentiate the attack traffic from the legitimate traffic for DDoS defense.

Even under a DDoS attack or a flash-crowd incident, it is important for a server to guarantee a Maximum Waiting Time (MWT) to its clients in addition to accurate waiting time estimates, since the MWT ensures that a client's request is accepted for the service within some finite time $T$. Unfortunately, existing fairness based (e.g., fair queueing) or proof-of-work based (e.g. computational puzzles) DDoS countermeasures fail to provide any MWT guarantee.

In this paper, we present RainCheck Filter (RCF), a DDoS mitigation primitive that enables a server to guarantee a MWT to clients when the server is overloaded. The core idea behind RCF is simple yet effective: RCF prioritizes clients' requests based on their waiting time (i.e., the time elapsed from the initial request) and rate-limit the number of requests per client, as if the server maintains an infinite queue in which the clients' requests are lined up waiting for the service. To simulate the infinite queue with a small physical buffer, in RCF the server sends the client a *raincheck*, a timestamped cryptographic token which not only tells the client when to retry but also serves as a proof of the client's priority level. In other words, an infinite virtual queue is simulated using rainchecks propagating over the network and stored at clients' buffer. Rainchecks are only valid for a limited time duration so that the server can efficiently rate limit each client and prevent raincheck reuse without keeping per-client state. RCF can be used to protect any critical resource, for example, as a middlebox in front of a flooded link or a server. Due to the rapid growth of DDoS

* This work was done while the authors were at Carnegie Mellon University.

attacks at the application layer [6], we mainly focus on applying RCF to protect critical resources at the server in this paper.

We prove that a RCF client can access the server within a finite time $T$ which is linear in the number of clients. Besides achieving strong guarantees, RCF is lightweight and extremely simple to deploy. Our implementation shows that RCF can operate at line rate, and requires minimal modification to servers and no modification to clients. We envision that RCF can work as a complementary defense to resolve server overload when detection-based schemes fail to block bots, and as a primary mediation for flash crowds.

Our main contributions are as follows:

- We present RCF, a lightweight DDoS mitigation primitive that helps legitimate clients obtain their fair share of the server's processing power by utilizing the network as an infinite virtual queue. RCF can mitigate flash crowds and be a last resort when it is ineffective or insufficient to separate the attack traffic from the legitimate traffic.
- We prove that RCF achieves MWT guarantees with only a small amount of state that is independent from the number of users, and that RCF does not require precise request scheduling, which none of the prior work can achieve.
- We thoroughly evaluate the performance and effectiveness of RCF using theoretical analysis, simulations, and an implementation. Our results confirm that in practice RCF not only guarantees bounded waiting time but also reduces variance in the waiting time. Such characteristics enable servers to provide reliable feedback to clients.
- We introduce a fully functional RCF Python module that can provide fine-grained (i.e. per-URL) protection to web developers with merely three additional lines of code. No modification is required on the client side.

## II. Problem Definition

Our primary goal is to provide Maximum Waiting Time (MWT) guarantees during server flooding, where the flood of incoming requests deplete the server's scarce resources (e.g., processor, disk I/O, or internal bandwidth.) As we are interested in the context of server flooding attacks, we assume that the network infrastructure, such as the link bandwidth, is well-provisioned.

A server may be flooded by either bots or legitimate clients, and the waiting time for legitimate clients may grow indefinitely. In a flash crowd, the server is swamped with requests from legitimate clients alone. In a bot-driven DDoS attack, the adversary directs compromised endhosts to overload the server with an overwhelming number of requests. In this paper, we make no assumption on the adversary's power and strategy. For example, a powerful and smart adversary can compromise the majority of endhosts and target one client to increase the client's waiting time.

We consider flooding by initial requests but not data following the requests, as the server can rate limit the requests and never accept requests at a rate higher than what it can support.

**Waiting time model.** Among various waiting time guarantees, we consider a strong notion called *maximum waiting time (MWT) guarantees* – a finite time $T$ within which a client's request is accepted for service [7].

From the server's perspective, the waiting time of a client request $c$ that is accepted after $r_c$ times of retries is: $T(c) = T(c, r_c) - T(c, 0)$, where $T(c, i)$ is the time at which the server sees the $i^{th}$ retry by client $c$, and the $0^{th}$ retry represents the original request. Similarly, the waiting time from a client's perspective, $T'(c)$, is the elapsed time from the time of client's initial request to the time that client receives the acceptance response from the server after $r_c$ retries: $T'(c) = T(c) + RTT(c) + \textit{process}(c)$, where *RTT(·)* and *process(·)* indicate the round trip time and the server's request processing time, respectively. Assuming that *RTT(·)* and *process(·)* are bounded, a bounded $T(c)$ implies a bounded $T'(c)$. Hence, without loss of generality we consider only the waiting time observed by the server.

**Server model.** When a request arrives, the server first performs some operations at line rate, such as replay detection. Since the incoming request rate, $R_{in}$, is bounded by the network line rate, the server can process every request before adding it to a queue. The queue is kept in fast memory such that enqueue and dequeue operations can be done at line rate as well. The server's processing rate, $R_s$, is limited by bottleneck operations (e.g., database query).

**Client model.** We consider a network of $N$ clients, consisting of $Z$ *compromised* and $N - Z$ *legitimate* clients, where $Z$ and $N$ are bounded but may be unknown to the server. Typically, compromised clients are controlled by an adversary via malware, while legitimate clients are controlled by their human users.

We assume that each client has a unique and unforgeable ID but make no assumption on how IDs are defined. Web developers can flexibly choose their identification methods, such as login, CAPTCHA, single-sign-on, IP address, or a combination of multiple factors. For example, a member-based web service can use login credentials as IDs to mitigate flash crowds. We discuss possible choices of client IDs and compare them in Appendix A.

**Desired properties.** A practical system should be efficient, immediately deployable, and usable. These desired properties are translated into the following design requirements:

- **MWT guarantees:** The DDoS-limiting primitive should bound the waiting time of a legitimate client,

and the bound should be independent of other clients' strategies.

- **Minimal overhead for both clients and servers:** The DDoS-limiting primitive should incur minimal overhead for both servers and clients, thereby avoiding the increase of the attack surface. In particular, the primitive should avoid per-request or per-client state on a server.
- **No modification to clients and the network infrastructure:** To support immediate deployment, the system should be easy to adopt on the server side and requires no modification to clients and the network infrastructure.
- **Accurate feedback:** The server's estimate of a client's waiting time should be within a reasonable error margin of the actual waiting time in order to increase users' willingness to wait.

## III. Raincheck Filter

Our core observation about RCF is that Maximum Waiting Time (MWT) guarantees can be achieved if the server keeps a large queue of size $N$, where $N$ is the number of clients. We call this an ideal buffer because in reality we would like to avoid keeping per-client state. RCF simulates the ideal buffer using a *realistic buffer* whose size is much smaller than $N$ by *leveraging the network as an infinite virtual queue*. Such a simulation is achieved through the exchange of a special type of message called *raincheck* between the client and the server.

We first present a simple approach using an ideal buffer and discuss its fundamental properties to achieve MWT guarantees. We then present RCF, which satisfies the fundamental properties, but with a realistic buffer. The notation is summarized below.

| | |
|---|---|
| $c$ | Client ID |
| $\rho_c$ | Client $c$'s raincheck |
| $N$ | Total number of clients |
| $R_s$ | Server's request processing rate |
| $Z$ | Number of compromised clients |
| $L$ | Server queue length ($L \ll N$) |
| $\Delta$ | Raincheck expiration period |
| $\Delta_{pause}$ | Pause time before resend |
| $t$ | Current time |
| $t_{start}$ | Start of the lifetime |
| $t_{end}$ | End of the lifetime |

Table I: Notation

### A. MWT Guarantees Using an Ideal Buffer

Using a buffer of size $N$, we can achieve MWT guarantees as follows. The buffer is modeled as a FIFO queue that enqueues incoming requests. By limiting each client to have no more than one request in the queue, we ensure that the buffer never overflows. Since the server can process $R_s$ requests per time unit, the waiting time is bounded by $\frac{N}{R_s}$.

This approach adopts a simple rate-limiting policy (i.e., one request per client in the queue) as well as a request-ranking policy that orders requests by their age, or the time during which a request has stayed in the buffer. The server processes the request with the lowest rank (i.e., the oldest request) first.

We observe that the request-ranking policy presents two properties that lead to MWT guarantees in this ideal case: (1) The initial rank of each request is bounded, and (2) the rank decreases over time. In Section IV-A, we generalize this observation and present a theorem that we use to prove RCF's MWT guarantee.

### B. RCF Design

With a buffer of size $L \ll N$, a flooded server has to discard most of the requests, making it difficult for the server to treat each client fairly and to bound the waiting time. To address this challenge, RCF leverages the network as an infinite virtual queue from which the server can retrieve the knowledge of previously dropped requests.

RCF is designed as a generic primitive that can be applied at different protocol layers and granularities. In this section, we describe high-level overview of the RCF protocol design, and present in Section VI-B the implementation details of RCF that supports per-URL protection at the application layer.

**Overview.** Figure 1 illustrates how RCF works on an overloaded server. The server can (1) accept, (2) reject, or (3) postpone an incoming request, which could be either raincheck-carrying or raincheck-absent. When the server needs to *postpone* a request, the server asks the client to revisit at a later time by issuing a raincheck to the client. Note that a raincheck-carrying request can be postponed again, in which case the client obtains an updated raincheck. To manage rainchecks, RCF implements two core components on the server side—*raincheck issuance* and *raincheck validation*—both of which operate at line rate.

The raincheck validation component checks a raincheck's validity using the server's secret key. Requests with an invalid (e.g., expired) raincheck are rejected, while requests with a valid raincheck are added to a priority queue of length $L$, in which a request that has waited longer gets higher priority.

For each valid yet dropped request, the raincheck issuance component constructs a raincheck using a server's secret key and returns the raincheck to the client. Rainchecks are protected using Message Authentication Codes (MACs) to prevent forgery, tampering, or sharing among multiple clients. The client can resend its request with the returned raincheck as a proof of the waiting time.

Raincheck-absent requests are forwarded to the raincheck issuance component directly (rather than being assigned the lowest priority) for two reasons: (1) To prevent the server from queuing requests with spoofed IP addresses (the client must return with the raincheck in order to use the server's service), and (2) to ensure bounded waiting time. Otherwise, a raincheck-absent request can be stuck in the queue forever when raincheck-carrying requests arrive at the same speed as the server's processing rate.

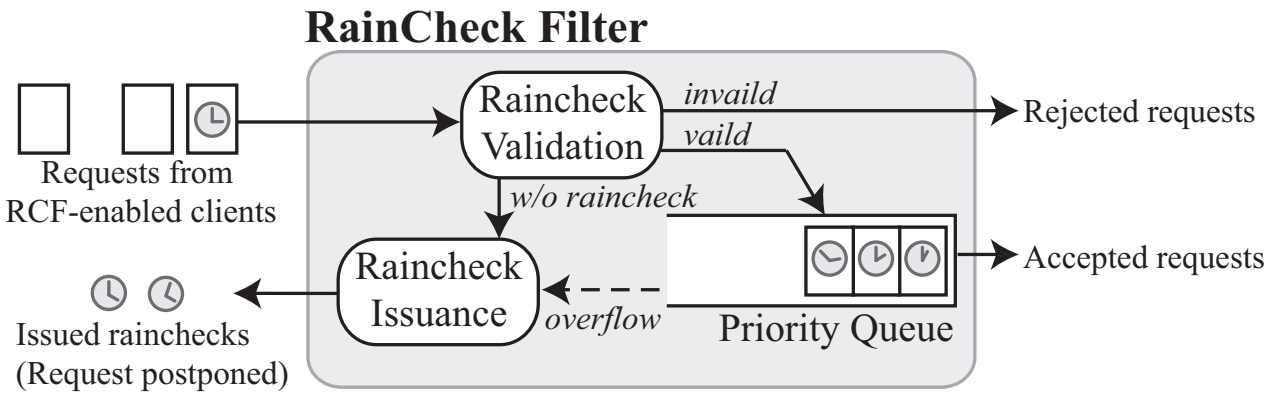

Figure 1: RCF overview.

One major challenge is how to prevent double-spending of rainchecks without keeping per-client state. To address this challenge, we impose on every raincheck a valid lifetime from $t_{start} = t + \Delta_{pause}$ to $t_{end} = t_{start} + \Delta$. Consequently, the server only needs to remember *accepted requests* for $\Delta_{pause} + \Delta$ period of time since any other raincheck of a client will expire after that. The server can adjust $\Delta_{pause}$ and $\Delta$ to strike a balance between communication and storage overhead.

**Raincheck message format.** A raincheck contains a MAC that protects client ID $c$, timestamp $ts$, and lifetime $[t_{start}, t_{end})$, all computed with the server's secret key $k$ such that an adversary cannot tamper with or forge a raincheck:

$$\rho_c = m \| MAC_k(m), \text{ where } m = c \| ts \| t_{start} \| t_{end}. \quad (1)$$

A unique client ID $c$ is included to enable rate limiting based on source identities and to prevent two clients from sharing their rainchecks. Since each MAC is computed using the server's secret key, only the server can correctly create and validate the raincheck. If the server enables more than one instance of RCF, each RCF should use a different secret key.

### C. Server Description

**Raincheck issuance.** A raincheck issued or renewed at time $t$ is valid from $t_{start} = t + \Delta_{pause}$ to $t_{end} = t_{start} + \Delta$, where $\Delta_{pause}$ is a small amount of minimum time that the client has to wait before resending. Moreover, when a raincheck-absent request arrives, the server drops the request directly and returns to the client $c$ a raincheck in which the timestamp is the current time. When a queue overflows, a raincheck-carrying request is dropped and directed to the raincheck issuance component for renewal but the timestamp stays the same as the one in the old raincheck.

Since the server issues a raincheck to every dropped request, a client can have multiple valid rainchecks concurrently. However, having multiple valid rainchecks provides no additional benefits to the client, because the server (or its raincheck validation component) limits the rate of accepted rainchecks per client as described below.

**Raincheck validation.** For efficient double-spending prevention and rate limiting, the server keeps a set $Accepted$ that contains requests that were accepted during time $[t - (\Delta + \Delta_{pause}), t)$. We denote by $Accepted(c)$ whether a client $c$'s request is in the set. Similarly, we denote by $Buffered(c)$ whether $c$'s request is currently buffered in the queue. Size of these records are irrelevant to the total clients and can be implementd efficiently using Bloom Filter variants [8], [9]. A raincheck is valid if all of the following conditions hold:

1. **Lifetime:** $t_{start} \leq cur_time < t_{end}$.
2. **No duplicate:** The same raincheck cannot be reused more than once
3. **Limited client request rate:** Only one request is allowed per client in any interval $\Delta$. That is, a client's request is accepted only if $Accepted(c) = False$ and $Buffered(c) = False$. This condition can be easily extended to allowing multiple requests per client in a time interval, which is useful when one instance of RCF is deployed to protect multiple critical resources.
4. **Integrity:** the MAC is verified correctly.

The first three conditions ensure that once a client's request is accepted, all other rainchecks that he possesses become invalid.

Valid requests are added to the priority queue and ranked unambiguously based on their timestamps.[1] If $L$ requests are already queued, this component dequeues the lowest priority request, issues a raincheck, and returns it to the client.

### D. Client Description

Figure 2 illustrates the client-server interaction in the RCF protocol. The client initiates the protocol by sending a raincheck-absent request. The server returns to the client a raincheck that expires after $\Delta$ in the future.

Before the raincheck expires, the client resends the raincheck-carrying request, and if the server is still busy, the client obtains a renewed raincheck with an extended lifetime. The client keeps resending until the request is accepted. We prove in Section IV that a client following this resend strategy will be able to access the server after a bounded delay.

A greedy client may attempt to reduce the waiting time by resending the request as quickly as possible. However, our protocol guarantees that the greedy client gains no benefit if he sends at a rate faster than $\frac{R_s}{L}$, because the buffer keeps only one copy of the request for each client. Moreover, the server explicitly specifies in the raincheck how long the client has to wait before retrying (i.e., setting $t_{start}$ in the raincheck's lifetime) to further restrain greedy clients and to minimize communication overhead.

[1] A trade-off exists between the timestamp granularity and the waiting time bound. When at most $v$ requests are allowed with the same timestamp value, the waiting time bound is increased by $\frac{v}{R_s}$.

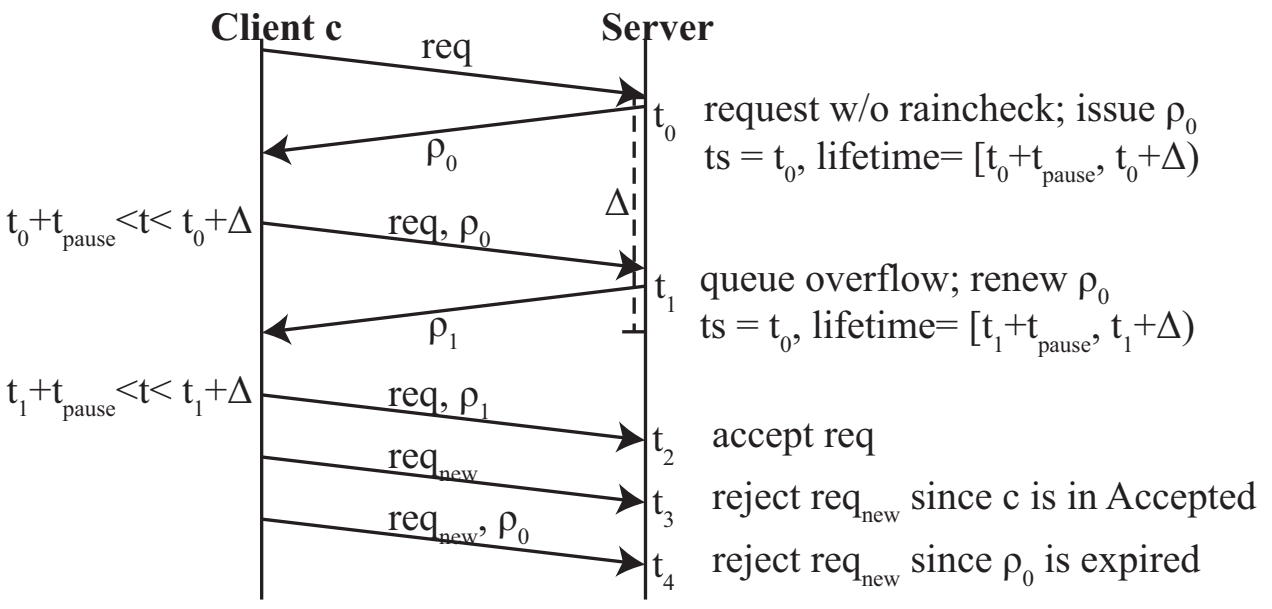

Figure 2: Client-server interaction example.

### *E. Configurations and Further Improvements*

To build a practical and deployable system, we have investigated the details of RCF's functionalities, corner cases, and the actual parameter configurations for MWT guarantees (e.g., minimum number of requests for various application scenarios, estimated number of requests that the bottleneck server can process given varying processing time, $\Delta$ configuration to satisfy performance requirements, etc.). We highlight the core ideas in this section, and leave the detailed description in the appendix.

**1. Accurate waiting time estimation.** Providing users with feedback of their expected waiting time can help increase their willingness to wait [3], [4]. While the waiting time bound $T$ can serve as a loose estimate of the actual waiting time, a better estimate that incorporates the current client status is desirable. We design and analyze a *rank estimation* algorithm that allows the server to estimate any client $c$'s rank (i.e., position in the virtual queue) at time $t$ without keeping per-client state. Extending probabilistic counting algorithms [10], our algorithm described in Appendix B refines the estimate by taking into account the number of clients that did not renew their rainchecks, and the server informs the client the estimated rank by piggybacking it on the raincheck.

**2. Balancing load distribution.** To avoid sudden increase in the bandwidth loads in the rare case when all clients concurrently retry, it is desirable to distribute the bandwidth loads caused by raincheck renewals. We propose a hybrid scheme that combines RCF with a coarse-grained scheduling for balancing load distribution in Appendix C. Similar to the original RCF, this hybrid scheme requires each client to renew its raincheck periodically. The novel improvement is the assignment of the coarse time interval such that all the requests from the same client always fall in the same time interval, reducing the overhead for duplicate detection and rate limiting.

**3. Accommodating multiple rainchecks.** There are applications where multiple rainchecks may be needed per client. For example, if RCF is applied to the HTTP protocol and each raincheck allows one HTTP request, multiple rainchecks would be needed to load a single webpage.[2] Getting a raincheck for each request sequentially would significantly increase the waiting time of the client. Our implementation in Section VI-B resolves this issue by enabling one RCF instance per resource-consuming URL, which likely accounts for a small set of all the HTTP requests for a webpage, such that the client can use multiple rainchecks in parallel. We also explore in Appendix D the effects of using an extended rate limit mechanism that allows a sender to have multiple requests in one virtual queue.

## IV. Analysis

### *A. Waiting Time Guarantees*

To prove that RCF guarantees MWT, we first show two properties of a rank function that imply MWT, and then that RCF satisfies these two properties.

Let $t_c^{ini}$ be the time at which the server sees client $c$'s first trial and $t_c^{acc}$ be the time at which the sever accepts $c$'s request/retry. We denote by $rank(c,t)$ client $c$'s priority of service (e.g., the position in the ideal queue) at time $t$, and $rank(c,t)$ is defined only for $t_c^{ini} \leq t < t_c^{acc}$. Client $c$ is served immediately at time $t$ when $rank(c,t) = 0$.

The rank function in RCF can be formulated as $rank(c,t) \triangleq |\{c'|c' <_t c\}|$, where $c_1 <_t c_2$ means $c_1$ has a lower rank than $c_2$ at time $t$ (i.e. at time $t$ client $c_1$ has a valid raincheck whose timestamp is smaller than any of client $c_2$'s valid raincheck.) When $rank(c,t) < L$ (i.e., server queue length), $c$'s request will be accepted.

*Theorem 1: Properties of a rank function ensuring MWT.* There is a bound $T$ such that for all $c$, $t_c^{acc} - t_c^{ini} \leq T$, and $T$ is independent of the attacker's power or strategy if the rank function satisfies the two conditions:

1. The initial rank of each client is bounded: $rank(c, t_c^{ini}) \leq B$ for all $c$, and $B$ is adversary-independent.
2. The rank of each client decreases over time: $\exists\ \delta > 0$ and $\gamma > 0$ such that $rank(c, t-\delta) - rank(c,t) \geq \gamma > 0$ for all $c$ and $t$, and $\delta$ and $\gamma$ are adversary-independent.

**Sketch of Proof:** Since it takes at most $\frac{\delta B}{\gamma}$ time to reduce a client's rank to zero, the waiting time is bounded: $T \leq \frac{\delta B}{\gamma}$. That is, the server guarantees MWT for any ranking function that satisfies the above two conditions. ■

*Theorem 2: RCF guarantees MWT.* RCF guarantees that a legitimate client will be served in a finite time $T$, regardless of how other (both legitimate or compromised) clients behave, and $T$ is linear in the total number of clients.

[2]A a typical webpage requires multiple HTTP requests, and modern browsers support concurrent HTTP connections (e.g., Firefox allows 15 concurrent connections).

**Sketch of Proof:** In RCF, the initial rank is bounded by $N$, the number of clients in the network. Between $c$'s $i$-th and $i+1$-th retries that are at least $\frac{L}{R_s}$ time apart, the server either accepts $c$'s request or accepts $L$ requests from the more privileged clients. Also, RCF ensures that once a client's request is accepted, all its rainchecks become invalid. Therefore, $rank(c,t)$ decreases by $L$ for every retry, which means after at most $\lceil \frac{N}{L} \rceil$ attempts the server will accept the request. Also, as specified in Section III, a legitimate client resends its request at a frequency $f$ such that $\frac{1}{\Delta+\Delta_{pause}} \leq f \leq \min\{\frac{R_s}{L}, \frac{1}{\Delta_{pause}}\}$. Hence, based on Theorem 1, the waiting time is bounded as follows:

$$T(C) \leq \left\lceil \frac{rank(C, t_c^{ini})}{L} \right\rceil / f \leq \left\lceil \frac{N}{L} \right\rceil (\Delta + \Delta_{pause}). \quad (2)$$

If the RTT is not negligible compared to $\Delta + \Delta_{pause}$, the bound should be revised to $\lceil \frac{N}{L} \rceil (\Delta + \Delta_{pause} + RTT)$ to compensate the delay. ■

The upper bound represents the worst case scenario where a strong attacker who knows the client's request sending schedule and controls every host except the victim client and server. In the presence of a realistic attacker not targeting a specific client, the actual waiting time can be much lower than the upper bound, because bots will keep consuming their high-ranked rainchecks while the legitimate client waits silently. This demonstrates a strength of RCF: the client can increase its priority by simply waiting, in contrast to prior work where the client has to "work", such as solving computational puzzles.

### B. Overhead Analysis

RCF incurs low computational, communication, and storage overhead for both servers and clients. Clients are not required to perform any additional computation. A client simply keeps the most recent raincheck for each protected resource, and renews the raincheck roughly every $\Delta + \Delta_{pause}$ time period.

Our implementation uses a 32-byte raincheck format: 32-bit client ID, 64-bit microsecond-level timestamp, 32-bit lifetime of the raincheck as an offset from the timestamp, and 128-bit MAC. Given this size, the server's computational overhead is minimal. For each generated raincheck, the server has to perform one MAC generation, and for each received raincheck, the server performs one MAC verification. With an efficient MAC function, rainchecks can be generated and verified at line rate. For example, it takes only 61 cycles (22ns) to compute a 128-bit CBC-MAC using AES on Intel i5-4430S that supports the AES-NI instruction set. Our implementation in Section VI confirms that enabling RCF does not degrade the service throughput.

RCF avoids keeping state for all clients at the cost of sending rainchecks, but RCF only keeps the "recently" (not every) accepted clients to support efficient validations (e.g., rate-limiting, duplicate detection) using expiration time. A server can further adjust the raincheck expiration period to strike a balance between communication and storage overhead.

RCF only incurs a small overhead to communication between servers and clients. For instance, when applying RCF to HTTP applications, the size of the HTTP header (where a raincheck would be stored) increases by less than 5%—the size of typical HTTP request/response headers are 700-800 bytes [11] and a raincheck is 32 bytes. Moreover, RCF explicitly specifies how long clients have to wait before retrying and thus incurs small overhead compared with the case where greedy clients or bots aggressively resend requests.

In RCF, initial requests without a raincheck are rejected, and a raincheck is issued and returned to the client. This initial rejection adds resiliency against IP spoofing attacks, but adds slight latency to serve the request. To minimize the initial latency, RCF can be dynamically enabled and disabled by the server. During peacetime (e.g., server utilization less than 70%), RCF remains inactive so that incoming new requests are served by the server. When the server utilization increases beyond the threshold, the RCF becomes active and start issuing rainchecks to incoming new requests.

### C. Security Benefits

We design RCF to avoid expanding the attack surface: Rainchecks are protected against forgery with Message Authentication Codes that can be generated and validated at line rate. RCF prevents traffic amplification attacks as rainchecks are smaller than typical HTTP request/response headers. Also, RCF is secure against source spoofing and other misuses (e.g., raincheck reuse, accumulation, or sharing), and prevents compromised or greedy clients from gaining an advantage over legitimate clients.

## V. Evaluation

To validate that RCF effectively simulates an infinite buffer and thereby enables bounded waiting time with low variance in the presence of flash crowds or DDoS attacks, we evaluate RCF using the NS-3 simulator. Our large-scale simulation measures the waiting time of legitimate users in two cases. Section V-A describes a flash-crowd case where a large number of legitimate users simultaneously try to access a server within a short-time period. Section V-B shows a DDoS attack case where a server is flooded by bots. For both cases, we compare results among (1) RCF, (2) a traditional client-server model with no protection, and (3) a computational puzzle [12].

### A. Flash-Crowd Effect

We consider 100,000 legitimate clients, and a server that can buffer up to 200 requests and process requests at a rate following an exponential distribution with an average of 5 ms. Every client makes one request where the initial request time is uniformly distributed across a 100-second interval, and hence, the server experiences on average 1,000 incoming requests per second, five times the rate of its average request processing capacity. Next we briefly describe the server and client models to simulate the flash-crowd effect.

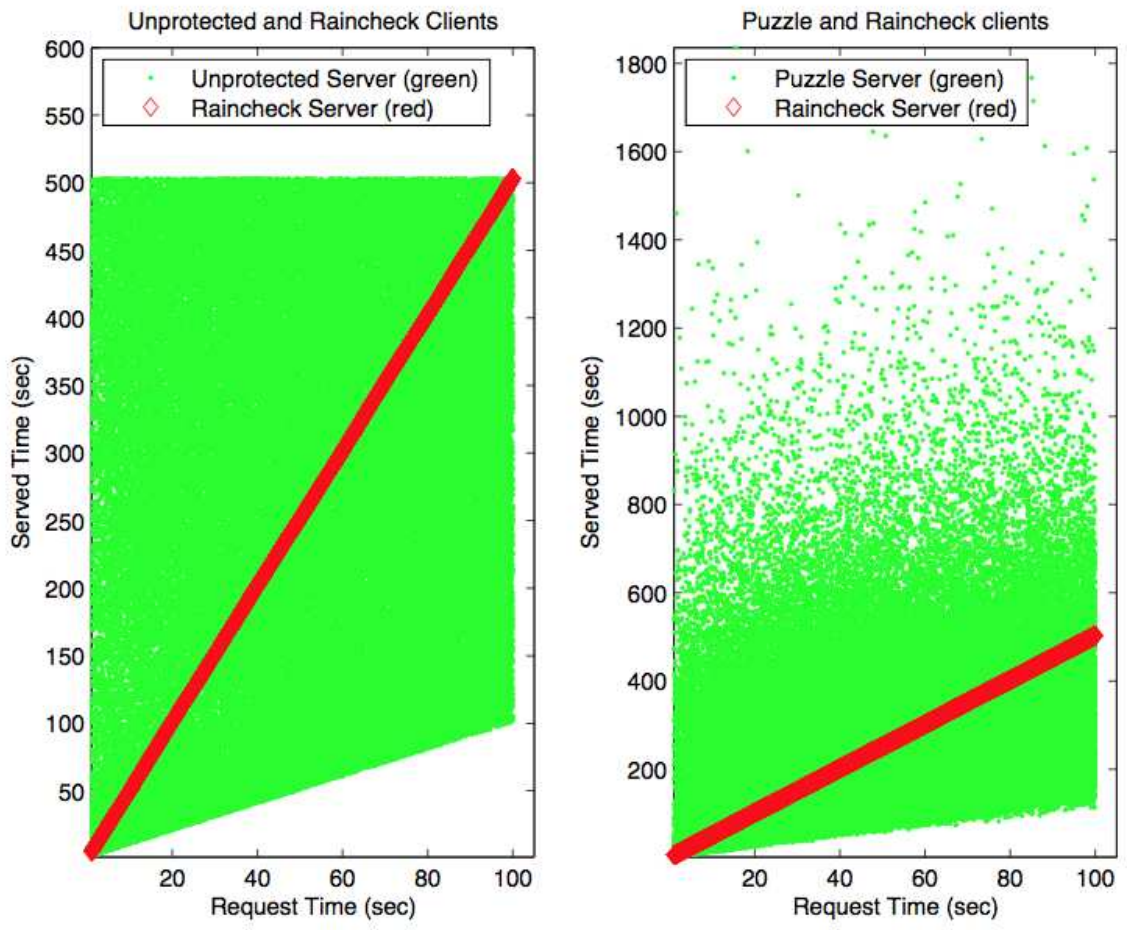

Figure 3: Scatter plots of the initial request time vs served time.

**Computational puzzles.** We model a client that solves a computational puzzle before sending a request [12]. We model the puzzle server with a priority queue which uses the puzzle-level as the priority metric. For requests with the same priority level, the server processes them based on their arrival order.

To send an initial request, the client solves a level-1 puzzle. If the request with level-$n$ is denied, the client solves a level-$(n+1)$ puzzle and resubmits its request. The puzzle computation time determines the delay that a client waits before submitting its request. The delay associated with a $l$-level puzzle is derived from the geometric distribution $0.01$geometric$(2^{-l})$, as the number of computed hashes before successfully finding a length-$l$ hash is geometrically distributed.

**Unprotected.** The unprotected server implements a standard FIFO queue and informs the client whenever its request is dropped. The client continues to resubmit requests until one of those request is processed by the server.

**Results.** The waiting times for the clients of the three models are shown as scatter plots in Fig. 3. Each dot or diamond in the scatter plot shows the requested time (x-value) and the accepted time (y-value) for a client's request. A green dot on the left figure represents a request of an unprotected client while a green dot on the right figure represents a request of a puzzle client. In both figures, the red diamonds represent the requests of the raincheck clients. The solidly filled-in green trapezoid areas in both figures demonstrate the large variances of the waiting times for unprotected and puzzle clients: some lucky clients are served almost immediately, while most clients are unlucky and experience long waiting times.

On the other hand, RCF clients, marked by red diamonds, resemble a thin line indicating that RCF supports low variance such that the waiting time steadily increases as the number of requests that are yet to be served increases. Specifically, the ordering of the requests is well-preserved—requests generated at earlier times are served before the requests generated at later times. Although not shown in the figure, the scatter plot for the raincheck server is almost identical to that for an ideal server that has an infinite buffer.

### B. Flooding Attacks

We consider 10,000 legitimate clients and bots ranging from 10,000 to 200,000 to observe the relationship between the volume of attack traffic and the waiting times that the clients experience. In this experiment, every client makes one request and the request times are uniformly distributed across a 200 second interval, and the servers' capacity are identical to the simulation in V-A.

To simulate the DDoS attack case, the bots adopt the following strategies to flood the servers. The bot's request generation model follows a Poisson process with $\lambda = 1$.

**Attacker strategy against RCF.** Since a RCF server favors requests with earlier timestamps, simple flooding (i.e., sending new requests at a high rate) is ineffective. Instead, a bot saves all valid rainchecks and sends the one with the earliest timestamp for each flooding period. Only when the bot runs out of valid (unexpired) rainchecks, it sends a new request.

**Attacker strategy against puzzles.** Since the puzzle server favors requests with higher-level puzzles, a puzzle bot submits a request with the highest puzzle-level that it solves in a requests generation interval. In addition to shorten the puzzle computation time, bots collaborate with each other, and the collaborative puzzle solving time is modeled as $\frac{0.01}{|bots|}$*geometric$(2^{-l})$.

**Attacker strategy against unprotected.** A bot periodically submits a request to the server.

**Results.** The distribution of the waiting times of the 10,000 clients are shown as box plots in Fig. 4. Similar to the flash-crowd simulation, this simulation confirms that the maximum waiting times for the raincheck clients are much smaller than the unprotected and puzzle clients regardless of the number of bots. In addition, the simulation also confirms that the variance of the waiting times are smaller than the other two cases (shown by the height of the boxes). Although not drawn in the figure, the maximum waiting times for the raincheck clients are below the upper bound in Eq. 2.

The maximum waiting time varies significantly for the unprotected clients. The increase in variance becomes more dramatic as the number of bots increases. This is because the unprotected server treats every request equally and the bot's collective request submission rate is much higher than that of the clients, depriving clients' access to the server.

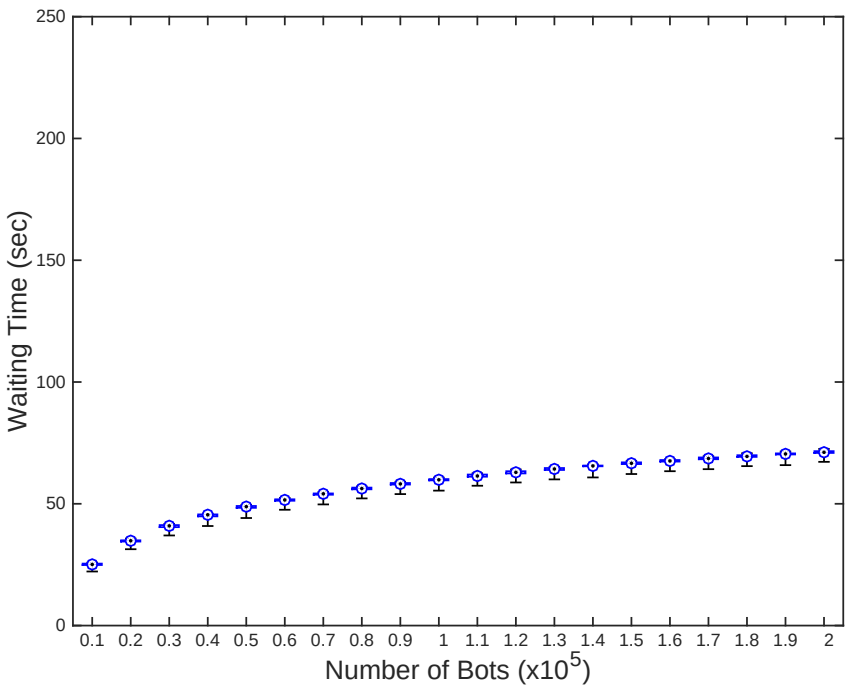

(a) RCF

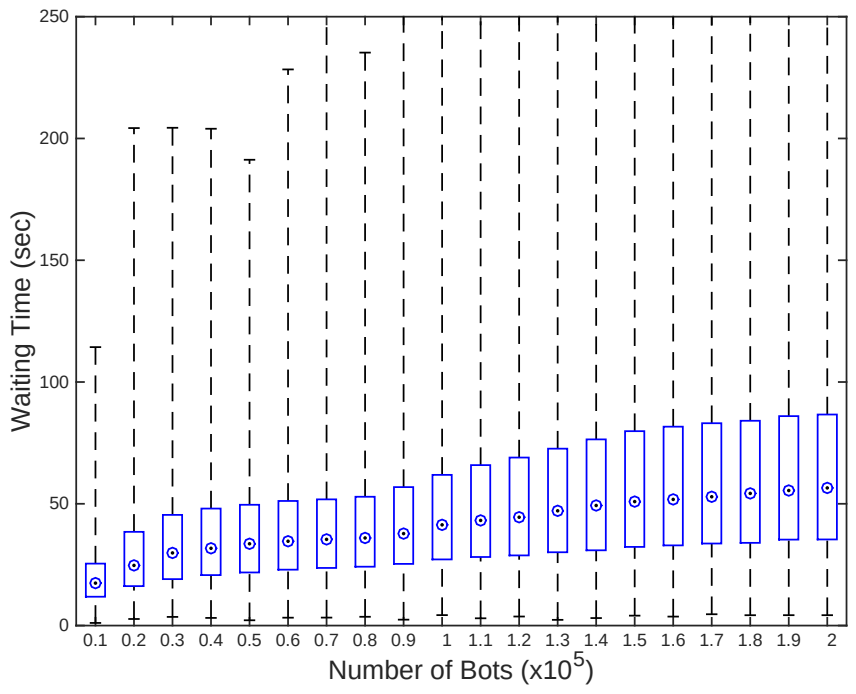

(b) Puzzles

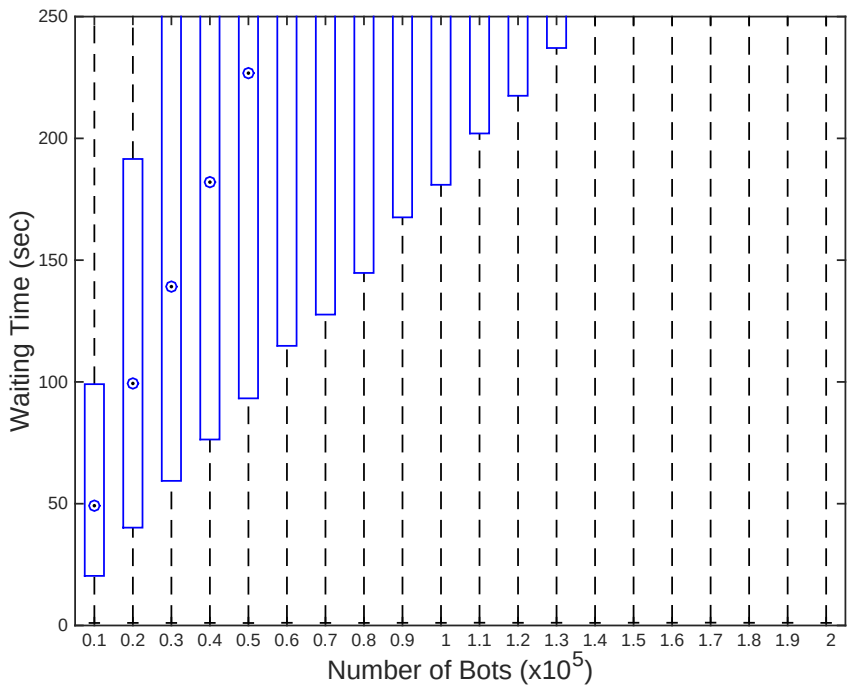

(c) Unprotected

Figure 4: Box plot of the waiting times for 10,000 clients under varying number of bots for the raincheck server, puzzle server, and unprotected server. Note that the mean waiting times for the unprotected server goes up to 800 sec (and thus outside the graph).

For the puzzle clients, while the median waiting time is comparable to that of the raincheck clients, the variance among the waiting times as well as the maximum waiting times are higher than that of the raincheck clients. This large variance is attributed to the amount of time needed to solve the puzzles.

## VI. Implementation

To demonstrate the efficiency and feasibility of RCF, we implement RCF (1) on a high-bandwidth testbed supporting line-rate raincheck validation and issuance, and (2) as a Python module that can be easily incorporated into HTTP servers.

### *A. DPDK Testbed Evaluation*

In this evaluation, we show that raincheck generation and validation can be done at line rate on a commodity machine and thus do not cause any additional bottleneck at the server.

**Testbed Setup.** We build a small testbed containing two machines—one machine to simulate a malicious traffic generator ($TG$) and the other to simulate a RCF. Two machines are directly connected to each other via a gigabit ethernet cable and operate on identical commodity PCs with the following hardware specification: 4-core Intel Core i5-4430S CPU (Haswell), 1,600 MHz Hynix DDR3 4GB memory, and one dual-port 10 GbE NIC (Intel 82599EB).

We use Intel Data Plane Development Kit [13] (DPDK) as the packet I/O engine on both machines to generate and process high-bandwidth traffic. Intel DPDK supports high-performance packet processing without involving the system kernel. DPDK avoids OS interrupts while packets are being transferred between the user space and the NIC's memory buffer to eliminate redundant memory copies in the kernel.

AES-CBC-MAC in RCF is implemented using the Intel AES-NI instruction set [14], a hardware accelerated cryptographic engine, to efficiently verify/issue rainchecks.

**Experiment Design.** $TG$ initiates the experiment by sending a request packet to RCF (Step 1). Upon receiving the request, RCF generates a raincheck and sends a packet with the raincheck back to $TG$ (Step 2). After this initial exchange, we start measuring the performance of RCF for raincheck validation and issuance in the following manner: $TG$ simply forwards the raincheck-carrying packet back to the server (Step 3). Then, RCF validates the raincheck, updates the raincheck with a new timestamp, and sends the updated raincheck back to $TG$ (Step 4). Steps 3 and 4 are repeated to measure the worst-case processing time of an incoming request.

For comparison, we establish the following baseline case: (1) $TG$ sends the initial requests to RCF, (2) RCF replies a packet with a dummy raincheck, (3) $TG$ immediately resends the received request back to the server, and (4) RCF simply forwards the packet back to $TG$ without verifying or updating the raincheck. Steps (3) and (4) are repeated for evaluation.

**Result.** For the baseline case, $TG$ sends raincheck-carrying requests to the server at a rate of 9.826 Gbps[3]. The same rate of 9.826 Gbps is achieved for the raincheck case. Even though RCF has to validate and issue a raincheck for each incoming request, we observed no degradation in throughput, confirming that the raincheck computation incurs a low overhead.

### *B. Raincheck over HTTP*

In this section, we present how RCF can be easily deployed to HTTP servers. Applying RCF to web servers can efficiently rate-limit requests that are compute-intensive (e.g., WolframAlpha) or IO-intensive (e.g., database queries) while ensuring per-client fair access.

Our implementation is a Python module that serves as an extension of Flask[4], a microframework for Python web development. This solution is practical in the real world because it requires no modification nor additional installation on the client side, and incurs minimal effort on server side. Specifically, our server-side implementation works as an intermediate level between the web server and the application, using Python's decorator feature such that developers can integrate RCF to existing server applications with little effort.

Listing 1 demonstrates the basic usage of the module. Text in bold is the additional code needed for applying the RainCheck module. Beside initialization (line 5), developers only have to add a line of code to mark the critical resource they want to protect (line 8). It is independent of applications being protected so that developers can easily turn RCF on and off even without any knowledge about the implementation details of applications.

The RainCheck module provides per-URL fine-grained protection, which is useful to flexibly apply RCF to critical resources only and leave noncritical resources (e.g., static content) freely accessible. Specifically, it allows different configurations and separated queues for different URLs, which is essential for web servers to provide various services.

[3]We cannot achieve the full capacity of 10 Gbps due to the bottleneck at the PCI that interconnects the NIC to the mainboard.

[4]http://flask.pocoo.org

```
1  from flask import Flask
2  app = Flask(__name__)
3
4  from raincheck import RainCheck
5  rc = RainCheck(queue_size=100, time_pause=1,
       time_interval=10, concurrency=4)
6
7  @app.route('/')
8  @rc.raincheck()
9  def index():
10     # compute-intensive or IO-intensive jobs
11     return resp
```

Listing 1: Basic usage of the RCF module.

**Implementation Details.** We fully implemented the RainCheck module described in Section III including issuance, validation, communication protocol, and ranking. The following are additional details and concerns of the implementation.

Rainchecks are carried through HTTP cookies, a common feature that the majority of clients support, so no additional modifications are needed on the client side. Cookies provide a simple mechanism allowing us to renew and expire rainchecks.

We use HTTP Refresh header to indicate how long a client should wait before coming back to renew its raincheck (i.e., no time syncronization is needed). The refresh time is uniformly random from $t_{start}$ to $t_{end} - t_{network_delay}$. Consequently, subsequent renewing requests will not cluster together regardless of clients' initial requesting pattern.

In addition to using the source IP as the client ID, we accept any entry in session storage[5] as the client ID, allowing developers to use login, CAPTCHA, single-sign-on, or other identification methods for the unique client ID generation, and addressing the issues with sharing the same public IP due to the prevalence of NAT.

The estimated current position of a client in the virtual queue is exposed to developers as a variable, such that developers can customize their web interfaces (e.g., show the estimated waiting time or error messages to clients).

The most computationally critical part of RCF is deriving message authentication codes. To integrate with the low overhead AES-NI implementation and be DoS-resilient, we use Python's ctypes library to invoke native functions and thus achieves comparable performance with Sec˙ VI-A.

**Raincheck over TCP vs. HTTP.** We now discuss the advantages and disadvantages of implementing RCF at the TCP layer and the HTTP layer.

Implementing RCF at the TCP layer may be more robust and efficient. It incurs less overhead as fewer layers are passed and requests can be dropped or sent back as soon as possible. When the bottleneck is out of application layer's sight (e.g., TCP flooding attack), implementation at the TCP layer can also protect more resources than that at the HTTP layer. Moreover, RCF can serve as a generic defense mechanism that protects all upper layer protocols and applications.

[5]In Flask, this is implemented on top of cryptographically-signed cookies.

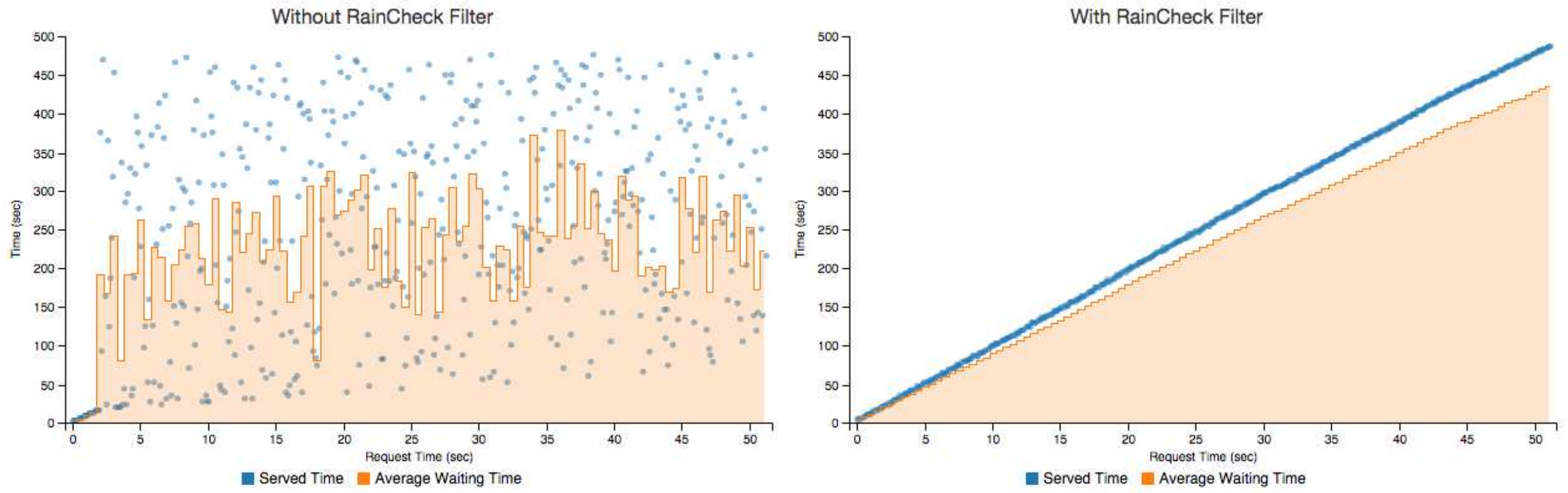

Figure 5: Scatter plots of the initial request time vs. served time and step charts of the average waiting time.

However, such a modification requires OS kernel changes or privileged module installation on both clients and servers which may be too intrusive and less favorable. It is also difficult to make changes in a widely used protocol and persuade people to update it at the same time. Furthermore, it may not be suitable to adopt a universal setting for various protocols and applications above TCP layer; some of them may even be unnecessary to protect.

On the other hand, implementing RCF at the HTTP layer is more flexible and transparent. Changes occur only on servers and developers have the full control to decide when to enable RCF, what resources to protect, and how to configure the queue size, waiting time, and other parameters. With the better knowledge of applications, RCF can be more accurately customized to a specific situation.

RCF can also be used to protect UDP packets, and RCF integration with UDP is relatively easier than with TCP, since UDP is connectionless. RCF can also be applied to protect other application requests, such as DNS queries.

**Parameter Selection.** We briefly describe here the rule of thumb in selecting parameters, particularly, determining reasonable ranges of $\Delta$ and $\Delta_{pause}$ within which changing these parameters does not affect much of the server's performance. Detailed mathematical analysis is included in Appendix E. The value of $\Delta$ can be determined by network delays and memory consumption: It should be long enough to tolerate the variance in network delay and short enough to keep small states of accepted clients. The value of $\Delta_{pause}$ can be determined by bandwidth consumption and rank feedback frequency: It should be long enough to prevent too much overhead on frequently raincheck renewal and short enough to update the ranking of clients promptly.

**Experiment.** We show the effectiveness of RCF over HTTP on a compute-intensive web server. We deployed the server on a m3.xlarge instance of Amazon EC2 service in Singapore, and used another EC2 instance in Tokyo to send requests. The configuration of the server is summarizes below.

| Server | without RCF | with RCF |
|---|---|---|
| Queue Size | 16 entries | |
| Concurrent Processing | 4 clients | |
| Computing Time per request | 1.5˜2.5 seconds | |
| $\Delta_{pause}$ | N/A | 1 second |
| $\Delta$ | N/A | 4 seconds |

For the unprotected server, we used a FIFO queue to buffer requests and to limit concurrent processing requests. (Otherwise, the server will be overloaded and all the requests will timeout.) The clients of the unprotected server retries in 1 to 5 seconds whenever the request fails until the request is accepted, mimicking the realistic scenario.

The server receives 10 requests per second uniformly from distinct clients for about 50 seconds. For the experimental purpose, we assume that clients already acquired unique IDs.

**Result.** Figure 5 shows the result of the server with and without RCF under the flash-crowd effect. The unprotected server suffers from high variances in the waiting time when the queue is full, whereas RCF server's waiting time is almost directly proportional to the initial request time. These consistent results validate our simulation's correctness and prove RCF's capabilities in the real world.

## VII. Related Work

One typical defense against service-level DDoS attacks aims to offer a fair chance of service access to clients. To avoid granting access to non-existing entities (e.g., via IP address spoofing) and to limit a client's attempt to gain advantage over others by masquerading multiple entities, DDoS defense mechanisms employ an interactive protocol requesting clients to present an evidence proving the clients' identity.

Computational puzzles [12], [15]–[18] demand clients to show their computational effort to get a service. Despite simple and stateless, they cause high overhead to legitimate clients while providing only weak probabilistic waiting time guarantee [7], which hinders their real world adoption.

CAPTCHAs [19] use a hard AI problem, which can be easily solved by most humans but not by machines (e.g., bots), to test the human presence behind a service request. CAPTCHAs have been widely adopted by many web-based applications to test human presence, and is also used to distinguish Flash Crowd and a DDoS attack [20]. However, advances in CAPTCHA breaking techniques [21]–[23] weaken its effectiveness as a DDoS defense tool. Furthermore, requirement for human interaction limits its applications.

We follow a line of thought of latency-based proof-of-work [24], where a server under a DoS attack prioritizes the requests of the clients who have waited long for the service. Crowcroft et al. proposed a mechanism to enforce passive delay on clients, slowing down the request rate. However, in contrast to raincheck, this mechanism needs per-client state at a server and does not provide any service access guarantee.

Various proposals aiming at a faster web [25]–[28] use a cryptographic credential (which is similar to TCP SYN cookie [29]) to reduce the number of round trips for the connection establishment. For example, TCP Fast Open (TFO) [25] speeds up successive TCP connections using a TFO cookie, a server-generated Message Authentication Code that proves the client's ownership of a source IP.

Technically, RCF creates credentials in a similar way to the aforementioned mechanisms. A key distinction is that each raincheck contains a fine-grained timestamp by which RCF performs admission control, guaranteeing a maximum waiting time for establishing a connection.

Queuing systems are heavily researched in computer science and operations research. Some mechanisms [30], [31] assign queues to aggregated requests by their origin. Among them, Lee et al. [31] proposed a mechanism that provides differential guarantees to the aggregates based on the observation that bot distribution is not uniform across domains. Stoica et al. [32] proposed a core stateless weighted fair queueing mechanism for fair network scheduling. Only edge routers maintain per-flow state while core routers use the labels inside packet headers created by edge routers to realize fair scheduling. However, queuing systems do not intend to offer nor can they provide precise waiting time guarantees to clients. Alternately, Gligor [7] proposed a scheme that provides per-client, maximum waiting time guarantees via precisely time-scheduled service-access tokens. Such scheduling requires conservative workload prediction for every single service and assumes all granted tokens would be used on time—which unavoidably leads to significant resource under-utilization.

Service replication via infrastructure outsourcing is a common practice for DDoS mitigation. However, many services, such as financial, government, and healthcare services, are hard to replicated/relocatable, distributed, or outsourced, e.g., for security and privacy reasons. Moreover, SSL-protected contents can only be served via a man-in-the-middle approach [33], which is highly undesirable from a security perspective. Traffic-scrubbing clouds are ineffective when it is difficult to differentiate malicious and legitimate clients. RCF outperforms prior works as it can be a simple yet practical solution to protect initial requests and guarantees access to a public service that cannot afford a server farm.

## VIII. Conclusion

Recent technology advances introduce unfortunate side effects: Internet users are becoming increasingly impatient. To increase users' willingness to wait, the waiting time should be kept small and with low variances, and users should be informed with accurate waiting time estimations. To this end, We propose RCF, a lightweight DDoS mitigation primitive that bounds the waiting time of a legitimate client. RCF achieves strong guarantees by leveraging the network as a virtual queue and ordering clients based on their arrival time, such that the resulting guarantees are close to the optimal case where the server has infinite memory. Since RCF focuses on bounding waiting time, it can work in conjunction with DDoS countermeasures that differentiate bots from legitimate clients to further strengthen the waiting time guarantees. Without RCF, there is little hope for legitimate clients to access the flooded server because the attacker who sends a large number of requests has huge advantage over legitimate clients. In contrast, with RCF, the server effectively provides legitimate clients with access guarantees in the presence of bot-driven DDoS attacks or flash crowds.

## Acknowledgements

This work has received funding from the European Research Council under the European Union's Seventh Framework Programme (FP7/2007-2013) / ERC grant agreement 617605, by the Ministry of Science and Technology, National Taiwan University, and Intel Corporation under Grants MOST-103-2218-E-002-034-MY2, MOST-103-2911-I-002-001, NTU-ICRP-104R7501, and NTU-ICRP-104R7501-1. We also gratefully acknowledge support by ETH Zurich and by NSF under award number CNS-1040801.

## APPENDIX

### A. Client ID Selection

RCF assumes each client has a unique and unforgeable ID and allows web developers to flexibly choose their identification methods, such as login, CAPTCHA, single sign-on, IP address, or a combination of multiple factors. Here we discuss possible choices of client IDs.

When using IP addresses as client IDs, it is assumed that an IP address represents the identity of a client, and RCF provides per-IP fairness, which effectively limits the sending rate of requests originating from each client. However, this assumption generally does not hold in practice as NATs are widely-deployed in the Internet, and all clients behind the same NAT have the same public IP address. Hence, if RCF treats each unique public IP address as one client, clients behind NATs will not be given a fair-share of the server's resource.

To provide fairness in this context, we present two approaches. One possible approach is to assign a unique ID that is independent of the IP address to each client. Such approach allows RCF to identify the clients that are behind the same NATs as each client would have different client ID. However, such an approach is susceptible to Sybil attacks, where a client creates multiple IDs to gain unfair advantage.

To mitigate the effects of Sybil attacks, the websites can require users to create user IDs and make the process of creating user IDs difficult to discourage from a malicious or greedy user from creating multiple IDs. To this end, CAPTCHA [19] or mobile SMS authentication can be used as part of the user ID creation.

Another approach to fair resource sharing is based on differential allocations based on prior connection history. The server can consider a fairness model where the allocation of server resources is proportional to the number of requests during peacetime (e.g., when the server is not flooded) [34]. Specifically, the server splits IP addresses into blocks and measures the number of requests served per address block during peacetime. Such information is used during an attack for a fair allocation of server resources based on the IP address space. However, this requires keeping state for each address block at the server.

### B. Waiting Time Estimation

Providing users with feedback of their expected waiting time can help increase their willingness to wait [3], [4]. While the waiting time bound $T$ can serve as a loose estimate of the actual waiting time, we desire a better estimate that incorporates the current client status. We design and analyze a *rank estimation* algorithm that allows the server to estimate any client $c$'s rank (i.e., its position in the virtual queue) at time $t$ without keeping per-client state. To achieve this goal, our rank estimation algorithm extends a probabilistic counting algorithm [10] so that the estimate can be refined over time by taking into account the number of clients that did not renew their rainchecks. The server can inform the client of the estimated rank by piggybacking it on the raincheck.

Efficient and accurate rank estimation is challenging: Counting the number of issued rainchecks with a smaller timestamp value than that of client $c$ is inefficient because it requires maintaining one counter for each client. Moreover, since a client can have multiple valid rainchecks at hand, the server should prevent counting the same client twice.

**Background of FM sketches.** To provide efficient and accurate estimation, our rank estimation algorithm extends a *probabilistic counting* algorithm proposed by Flajolet and Martin [10], which estimates the number of distinct items in a set using $O(\log N)$ memory, where $N$ is the number of distinct items. The probabilistic counting algorithm keeps a bit vector (referred to as an FM sketch) that is initialized to zeros, and uses a deterministic function to map an item to the $i$th bit with probability $1/2^i$. A bit is set to 1 if an item is mapped to that bit. Given an FM sketch $V$ the estimated number of distinct items is

$$\tilde{n} = 2^{lsb_0(V)}/0.77351,$$

where $lsb_0(V)$ is the lowest-order 0-bit position of $V$ (zero-based indexing). One can reduce the estimation error by averaging the results of multiple FM sketches with different index functions. The FM algorithm guarantees a bounded error such that

$$Pr[|\tilde{n} - n| < \epsilon N] > 1 - \delta$$

with $O(\frac{\log(2/\delta)}{\epsilon^2})$ number of sketches.

**Our rank estimation algorithm.** Before explaining how our algorithm works, we first define several notations. Let $U^t_{t-\Delta}(x)$ contain all clients obtaining at least one raincheck with a timestamp $\leq x$ during $[t-\Delta, t)$. Let $n^t_{t-\Delta}(x)$ be the size of $U^t_{t-\Delta}(x)$. Since a client in front of $c$ at time $t$ must be in $U^t_{t-\Delta}(TS(\rho_c))$, $rank(c, t) \leq n^t_{t-\Delta}(TS(\rho_c))$, where $\rho_c$ is client $c$'s raincheck that has the smallest timestamp value among all valid rainchecks at time $t$ and $TS$ stands for timestamp.

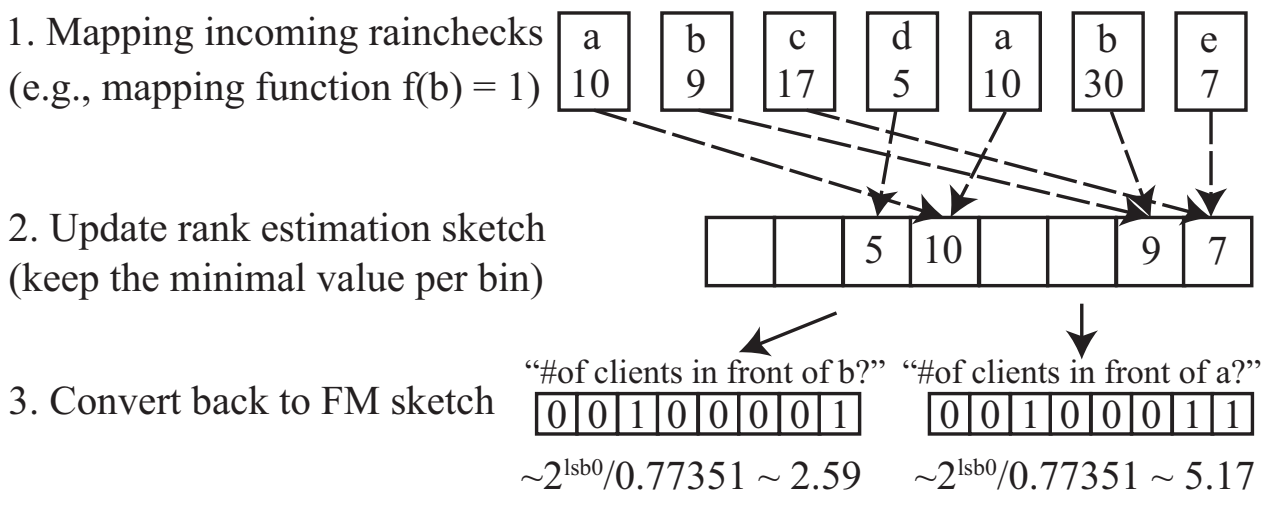

Figure 6: A rank estimation sketch example.

Estimating each client's rank separately using FM sketches requires $O(N \log N)$ memory. In contrast, our rank estimation algorithm is designed to answer queries such as "How many distinct items have a value lower than $x$?" for any $x$, while keeping only $O(\log N)$ number of items.

Specifically, the update and estimate operations in our rank estimation algorithm work as follows:

- **Update:** Instead of setting a bit to one, we store the item's value at the mapped position if the value is lower than the currently stored value at that position.
- **Estimate**: To estimate the number of items whose value is lower than $x$ for any $x$, we convert this rank estimation sketch back to an FM sketch where a 1-bit is set if the corresponding position has a value $< x$.

Figure 6 illustrates how the rank estimation sketch works and its relationship with FM sketches.

The core idea is that the rank estimation sketch ensures that the resulting FM sketch is the same as the one we can get by running the FM algorithm over items with a value lower than $x$. In our setting, items are rainchecks, and each raincheck has a value which is a timestamp. Two rainchecks are "identical" if they are from the same client. Hence, using this rank estimation algorithm, we can estimate $n^t_{t-\Delta}(x)$ for any $x$ given a time interval $[t-\Delta, t)$.

To estimate a client's rank at time $t_{cur}$ in the virtual queue, the server maintains a rank estimation sketch and resets it periodically every $\Delta$ time, such that the sketch accounts for intervals $[(i-1)\Delta, i\Delta)$ for all positive integers $i$. The rank $rank(c, t_{cur})$ can be approximated by the following:

$$rank(c, t_{cur}) \leq rank(c, i\Delta) \leq n^{i\Delta}_{(i-1)\Delta}(TS(\rho_c)),$$

where $i\Delta \leq t_{cur} < (i+1)\Delta$.

Based on the proof in Section IV-A, $rank(c, i\Delta) - rank(c, t_{cur}) \leq (t_{cur} - i\Delta)R_s$. Also, $n^{i\Delta}_{(i-1)\Delta}(TS(\rho_c)) - rank(c, i\Delta) \leq \Delta R_s$. Hence, with $O(\frac{\log(2/\delta)}{\epsilon^2})$ number of sketches, we ensure that the estimation error is less than $\epsilon N + (t_{cur} - (i-1)\Delta)R_s \leq \epsilon N + 2\Delta R_s$ with probability higher than $1-\delta$.

### C. Balancing Load Distribution

To avoid sudden increase in the bandwidth loads in the rare case when all clients retry concurrently, it is desirable to distribute the bandwidth loads caused by raincheck renewals. We propose a hybrid scheme that combines RCF with a coarse-grained scheduling for balancing load distribution. Similar to the original RCF, this hybrid scheme requires each client to renew its raincheck periodically. The novel improvement is the assignment of the coarse time interval such that all the requests from the same client always fall in the same time interval, thereby reducing the overhead for duplicate detection and rate limiting.

In particular, the server divides the time into non-overlapping time intervals $\delta_i = [i \cdot w, (i+1) \cdot w)$ for some constant $w$. A raincheck issued during $\delta_i$ is valid only during $\delta_j$ where $j \in [i+m_{min}, i+m_{max})$ for some required cooling period $m_{min}$ and expiration period $m_{max} = m_{min} + m_\Delta$. $m_{min}$, $m_{max}$, and $m_\Delta$ are positive integers. The integer value $j$ is derived such that

$$i + m_{min} \leq j < i + m_{max},$$
$$j \mod m_\Delta = PRF_k(cid) \mod m_\Delta,$$

where $k$ is a secret key and $PRF$ is a pseudorandom function. These equations have a unique solution.

This construction ensures that every raincheck-carrying request from the same client always comes back during the same time interval. Also, requests are renewed approximately every $m_\Delta$ time intervals. Therefore, the server only needs to keep track of the accepted requests during the current time interval, $\delta_i$, for duplicate detection and rate limiting. This hybrid scheme is easier to implement as the server does not need to maintain a sliding window of $\Delta$ as in the original RCF.

The server may want to update the secret key to increase randomness and minimize the risk of key exposure. Suppose the server would like to completely switch to a new key $k'$ from the beginning of $\delta_v$. To ensure a smooth transition, the server chooses to use either the old key or the new key from the beginning of $\delta_{v-m_{max}}$ on a per-request-basis according to the following criterion: if $j < v$ when computing $j$ using the old key, use the old key; otherwise use the new key.

This hybrid scheme trades flexibility for scalability. At one extreme where the PRF perfectly distributes the clients among $m_\Delta$ intervals, the overhead (e.g., storage or bandwidth consumed by raincheck renewal) is reduced by an order of $m_\Delta$. In particular, we obtain a MWT guarantee $\lceil \frac{N}{L} \rceil w$ when $w \geq \frac{L}{R_s}$. We omit this proof since it is similar to that in Section IV. At the other extreme where the PRF maps every client to the same time interval, it is degenerated to the original RCF with $\Delta = m_{max} \cdot w$, rendering a MWT bound $\lceil \frac{N}{L} \rceil m_{max} w$. Since any practical implementation of PRF should generate reasonably randomized outputs given that the key is kept secret, this hybrid scheme is expected to have much better scalability than the original RCF.

The server can adjust $m_{max}$ dynamically to distribute the load among $m_{max} - m_{min}$ intervals such that the server is slightly overloaded during each interval. The $m_{min}$ value should be large enough to ensure that requests can return to the server on time.

### D. Accommodating Multiple Rainchecks

There are applications where multiple rainchecks may be needed per client. For example, if RCF is applied to the HTTP protocol and each raincheck allows one HTTP request, multiple rainchecks would be needed to load a single webpage.[6] Getting a raincheck for each request sequentially would significantly increase the waiting time of the client. Our implementation in Section VI-B resolves this issue by enabling one RCF instance per resource-consuming URL, which likely account for a small set of all the HTTP requests for a webpage, such that the client can use multiple rainchecks in parallel. Here we explore the effects of using an extended rate limit mechanism that allows a sender to have multiple requests in the virtual queue.

[6] A a typical webpage requires multiple HTTP requests. Modern browsers support concurrent HTTP connections. For example, Firefox allows 15 concurrent HTTP connections.

RCF can be extended to accomodate *n*-rainchecks per client to support similar cases as above. When a client's raincheck is accepted, at most *n-1* additional rainchecks can be renewed or accepted for the same client during $\Delta$. That is, at any point in time, the total number of client's requests in the server should be less than or equal to *n*. RCF can also issue rainchecks to resource-consuming requests only, which likely account for a small set of all the HTTP requests for a webpage.

**Simulation: Effects of Supporting Multiple Rainchecks.**

The parameters that are used for the simulations in flooding attacks (Section V-B) are used. However, instead of submitting one request, each client submits five requests, and the amount of time that is needed to have all five requests accepted by the server is measured.

Client's request sending strategy is modified to handle multiple requests. A raincheck client sends all five requests simultaneously unless the server allows a smaller number of simultaneous requests. If the server allows a fewer number of simultaneous requests, the client initially submits the maximum number of requests that are allowed by the server, and for each accepted request, the remaining requests are submitted sequentially after waiting for $\Delta$. For a puzzle client, the client initially submits all five requests with the lowest-level puzzles. For each returned request, the client resubmits the request with a puzzle that is a level higher than the previous puzzle. For an unprotected client, the client submits all five requests simultaneously. Then for each returned request, the client immediately resubmits the request to the server.

Figure 7 summarizes the waiting times for the clients to get 5 requests accepted. The leftmost figure 7 shows the longest waiting times for the 10,000 raincheck clients when the server allows 3 and 7 simultaneous requests. Furthermore, a box plot showing the distribution of the waiting times for the 10,000 clients are shown for the case where the server allows 5 simultaneous requests per client. The figures in the middle and the right describe the distribution of the waiting times for the 10,000 clients for the puzzle and normal servers, respectively.

As expected, clients of all three models need to wait longer to get their 5 requests accepted by the servers. The primary reason for the longer waits are due to the increase in client's traffic—each client submits 5 requests. Moreover in the raincheck model, the volume of bots' traffic increases as well as bots take advantage of multiple simultaneous rainchecks that the server allows.

Even in this simulation where the waiting time is measured as the time to get all 5 requests accepted, the raincheck clients have the shortest maximum waiting time as well as the smallest variance in the waiting times.

The maximum waiting time for a raincheck client is the shortest when the raincheck server allows 5 simultaneous requests. The waiting time increases when the server allows 7 simultaneous requests because it allows bots to flood the server at the highest rate among the three cases. Although bots' flooding rate is the lowest, client's waiting time increases significantly when the server allows 3 simultaneous request. This is because the client's fourth and fifth requests can only be submitted after two of the first three requests are accepted; hence, the timestamps in the rainchecks for the last two requests are correspondingly late as many requests are already scheduled in the virtual queue.

The experiment suggests that if RCF were to be used for HTTP servers and clients, the waiting time that clients would experience depends on the number of simultaneous requests that are allowed by the server. Moreover, the experiment suggests that it is better for a server to allow a larger number of simultaneous requests per client than a smaller number of simultaneous requests compared to the number of requests that are actually needed to load the web-pages as the latter (allowing smaller number of simultaneous requests) may significantly increase the waiting time.

### *E. Parameter Configurations*

**Bounding and determining the request processing rate.** To provide MWT guarantees, the bottleneck (e.g., the server) is required to process at least $R_s$ requests per second. We discuss how to ensure a lower bound on $R_s$ under several application scenarios. We also discuss how to estimate $R_s$ when request processing time varies. While accurate estimation of $R_s$ (client's waiting time) can improve user experience, RCF's operations and guarantees do not depend on an accurate modeling of $R_s$.

When the server's access link is the bottleneck, RCF has to be installed in front of the access link, such as at a firewall or a load balancer. Since the bandwidth (BW) is the critical resource, raincheck can determine a lower bound of the request processing rate by $R_s \geq \min\{\frac{\text{downlink_BW}}{\text{max_req}}, \frac{\text{uplink_BW}}{\text{max_resp}}\}$, where max_req and max_resp are the maximum sizes of the request and response packets, respectively. A similar approach can be applied to the case that an intermediate network link is the bottleneck.

The attacker can also try to exhaust the CPU resource. Although the server has full knowledge about the type of service it offers, the time to complete a request may still be unpredictable beforehand. One possible solution is to consider a raincheck to be an explicit permission for accessing a unit of resources. Moreover, clients can wait longer to obtain higher level rainchecks that give permissions to access more resources. For example, one raincheck can represent 10k CPU cycles or 10ms of the server time. To adopt this modification, either (1) the client divides the task into smaller chunks so that each chunk can be processed in one unit of resources, or (2) the server terminates the process for the accepted request if it has run out of one unit of resources, and returns the necessary information such that the server can resume the process later, if possible. The server could also decline such computational-expensive requests during DDoS attacks (which is similar to a safe mode that only supports limited functionality). We leave it as future work to explore such resource allocation polices in the RCF framework.

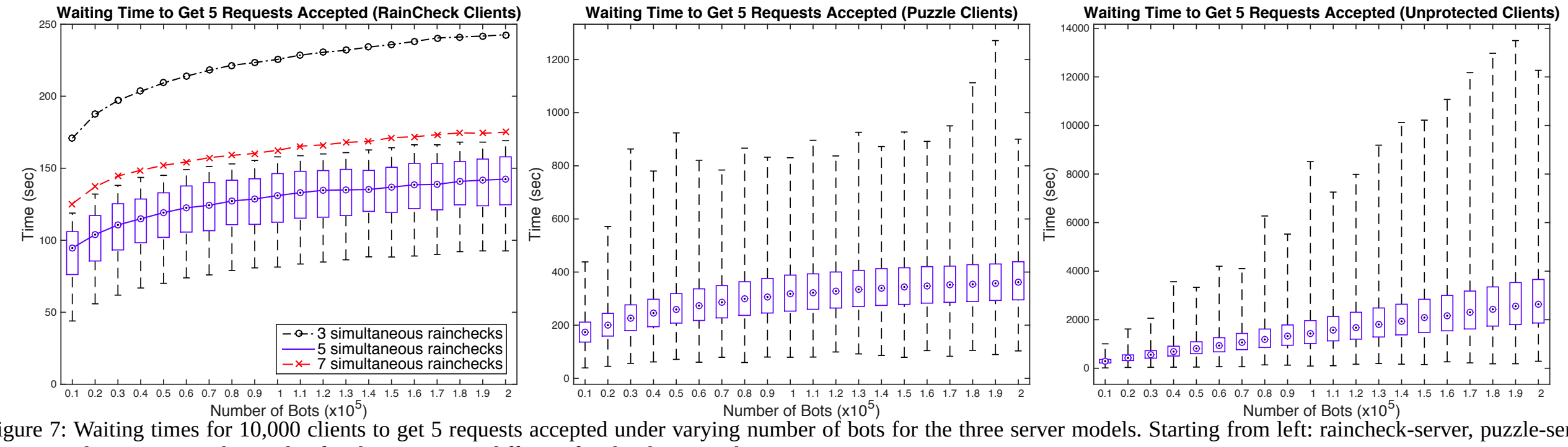

Figure 7: Waiting times for 10,000 clients to get 5 requests accepted under varying number of bots for the three server models. Starting from left: raincheck-server, puzzle-server, unprotected-server. Note that scales for the Y-axis are different for the three graphs

When the request completion time is predictable, the server can allocate resources to multiple request groups, each of which consists of the requests with similar completion time (e.g., based on their service types). Grouping similar requests not only improves the accuracy of waiting time estimation but also allows the server to apply RCF to resource-consuming requests only.

RCF can also help address memory exhaustion attacks such as TCP SYN flooding or slow HTTP attacks. Rainchecks can work like SYN cookies, which push state back to the client. In addition, by setting an explicit timeout on each buffered request, the request processing time is bounded too and thus RCF can be applied.

**Configuring $\Delta$.** Let $M$ be the size of the memory (in terms of the number of requests) available for RCF. Recall that $R_{in}$ is the incoming request rate, $R_s$ is the server's request processing rate, $\Delta$ the expiration period, and $N$ is the number of clients.

According to Section III, clients wishing to stay in the virtual queue must renew their rainchecks before they expire. Hence, the expiration period should be long enough to accommodate every client in the worst case. That is, $R_{in} \cdot \Delta \geq N$. On the other hand, the expiration period should be short enough to avoid keeping too much state at the server: $R_s \cdot \Delta \leq M$. These two constraints can serve as guidelines for RCF configuration.

Since the server typically has no control over $N$ and may be unable to immediately increase memory, we can set $\Delta$ to be $\frac{\tilde{N}}{R_{in}}$, where $\tilde{N}$ is the estimate of the number of clients. The server can obtain $\tilde{N}$ based on the recent history, and adjust $\Delta$ accordingly if the estimation changes. If the $\Delta$ value does not satisfy the second constraint, the server can either reduce the request processing rate to $R_s = \frac{M}{\Delta}$ or increase its memory size if possible. Note the reducing $R_s$ to $\frac{M}{\Delta}$ does not affect the MWT since $R_s$ is still higher than $\frac{L}{\Delta}$, which is required for obtaining the MWT bound in the worst-case scenario in Section IV.

We now briefly discuss how to set the parameters for the hybrid scheme described in Appendix C. In the hybrid scheme, the server similarly has to satisfy $R_{in} \cdot m_\Delta \cdot w \geq N$ and $R_s \cdot w \leq M$. Given the introduction of an additional parameter, $m_\Delta$, we can fulfill the second criterion first by setting $w = \frac{M}{R_s}$. We then set $m_\Delta = \frac{\tilde{N}}{R_{in} \cdot w}$.